\documentclass[journal=jacsat,manuscript=article]{achemso}

\usepackage[version=3]{mhchem} 
\usepackage[T1]{fontenc}       
\usepackage{xcolor}

\author{C. Godfrin}
\affiliation{CNRS Inst. NEEL, F-38000, France.}
\alsoaffiliation{Univ. Grenoble Alpes, Inst. NEEL, F-38000, France. }
\author{S. Thiele}
\affiliation{CNRS Inst. NEEL, F-38000, France.}
\alsoaffiliation{Univ. Grenoble Alpes, Inst. NEEL, F-38000, France. }
\author{K. Ferhat}
\affiliation{CNRS Inst. NEEL, F-38000, France.}
\alsoaffiliation{Univ. Grenoble Alpes, Inst. NEEL, F-38000, France. }
\author{S. Klyatskaya}
\affiliation{Institute of Nanotechnology (INT) Karlsruhe Institute of Technology (KIT) 76344 Eggenstein-Leopoldshafen, Germany}
\author{M. Ruben}
\affiliation{Institute of Nanotechnology (INT) Karlsruhe Institute of Technology (KIT) 76344 Eggenstein-Leopoldshafen, Germany}
\alsoaffiliation{Institut de Physique et Chimie des Mat\'eriaux de Strasbourg (IPCMS), CNRS-Universit\'e de Strasbourg 67034 Strasbourg, France.}
\author{W. Wernsdorfer}
\affiliation{Institute of Nanotechnology (INT) Karlsruhe Institute of Technology (KIT) 76344 Eggenstein-Leopoldshafen, Germany}
\alsoaffiliation{CNRS Inst. NEEL, F-38000, France.}
\alsoaffiliation{Univ. Grenoble Alpes, Inst. NEEL, F-38000, France. }
\author{F. Balestro}
\affiliation{CNRS Inst. NEEL, F-38000, France.}
\alsoaffiliation{Univ. Grenoble Alpes, Inst. NEEL, F-38000, France. }
\alsoaffiliation{Inst. Univ. de France 103 Blvd Saint-Michel 75005 Paris, France. }
\email{franck.balestro@neel.cnrs.fr}
\phone{+33 (0)476887915}

\title[An \textsf{achemso} demo]
  {Electrical single spin read-out using an exchange-coupled quantum dot}

\abbreviations{IR,NMR,UV}

\begin{document}


\begin{abstract}
We present a new way of continuously reading-out the state of a single electronic spin. Our detection scheme is based on an exchange interaction between the electronic spin and a nearby read-out quantum dot. The coupling between the two systems results in a spin-dependent conductance through the read-out dot and establishes an all electrical and non-destructive single spin detection. With conductance variations up to 4\% and  read-out fidelities greater than 99.5\%, this method represents an alternative to systems where spin to charge conversion cannot be implemented. Using a semi-classical approach, we present an asymetric exchange coupling model in good agreement with our experimental results.
\end{abstract}


Over the last 10 years, advances in nanofabrication and measurement technologies allowed for the read-out~\cite{Elzerman2004,Morello2010,Neumann2010,Vincent2012a} and manipulation~\cite{Jelezko2004,Koppens2006,Pioro2008,Koehl2011,Pla2012,Pla2013,Thiele2014} of single electronic and nuclear spins. Besides the opportunity of testing our understanding of quantum mechanics, these progresses are at the heart of recent developments towards potential applications in the field of nanospintronics~\cite{Wolf2001,Seneor2007}, molecular spintronics~\cite{Rocha2005,Bogani2008,Urdampilleta2011} and quantum information processing~\cite{Loss1998}. Among different concepts, systems integrating an all electrical spin detection benefit most from achievements of the microelectronic industry, but so far, they relied on the spin to charge conversion~\cite{Field1993,Morello2010,Meded2011} which required emptying the electron into a nearby reservoir. However, in devices where the energy of the spin system is much more negative than the Fermi energy of the leads, a different detection technique is mandatory. Here we present a general detection scheme based on a single electronic spin coupled to a read-out quantum dot. Because of the exchange coupling between the two systems, transport properties through the read-out quantum dot are spin dependent, enabling an electrical read-out which is non-destructive of the single electronic spin.

To implement this general detection scheme, we fabricated a single TbPc$_2$ molecule spin-transistor (Figure \ref{fig1}a) using the electromigration technique\cite{Park1999}. This spin transistor can be split into two coupled quantum systems :

(i) The 4f electrons of the terbium Tb$^{3+}$ ion possess an \textbf{electronic spin}. Its total spin S=3 and total orbital momentum L=3, originats from its [Xe]4f$^8$ electronic configuration. Because of a strong spin-orbit coupling, the total angular magnetic moment of the electronic spin is J=6. In addition, the two phthalocyanines (Pc) generate a ligand field that leads to an electron-spin ground state doublet $|\uparrow\rangle$ and $|\downarrow\rangle$ which is well isolated and have a uniaxial anisotropy axis perpendicular to the Pc-plane \cite{Ishikawa2003} (Figure~\ref{fig1}a). At finite magnetic fields, the degeneracy of the doublet is lifted, the electronic spin can reverse and emit a phonon \textit{via} a  direct relaxation process. 

(ii) the Pc ligands create a \textbf{read-out quantum dot}. The TbPc$_2$ has a spin S=$1/2$ delocalized over the two Pc ligands\cite{Vitali2008} which is close in energy to the Tb-4f states. Thus, delocalized $\pi$-electron system results in a quantum dot in the vicinity of the electronic spin carried by the Tb$^{3+}$ ion. This read-out quantum dot is tunnel-coupled to source and drain terminals to perform transport measurements. Furthermore, an overlap of this $\pi$-electron with the Tb$^{3+}$ 4f electrons gives rise to a strong exchange coupling  between the read-out quantum dot and the electronic spin, without affecting its magnetic properties, as demonstrated in the following.

\section{Experimental Results}

\begin{figure}
\begin{center}
\includegraphics[width=0.4\textwidth]{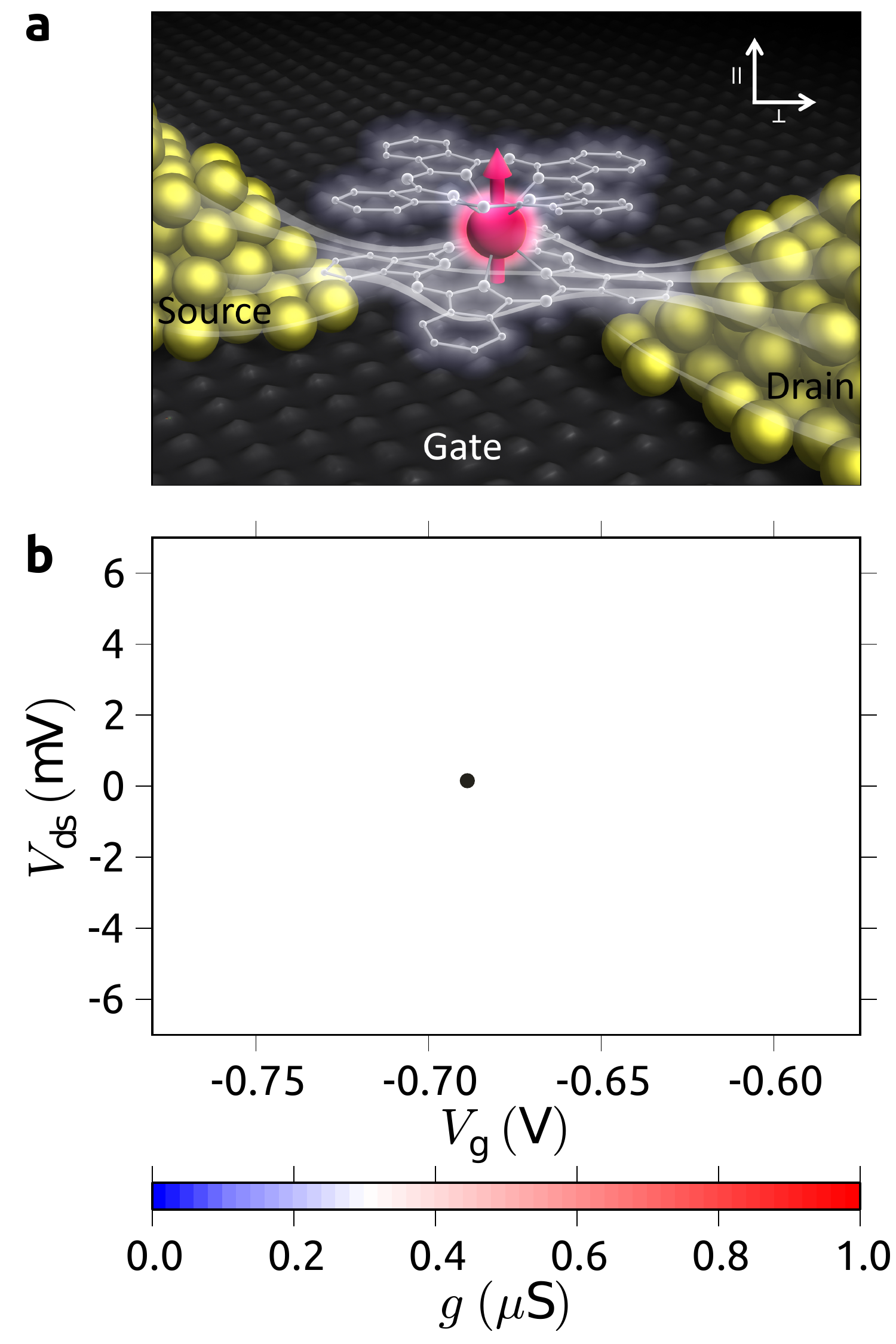}
\caption{\label{fig1}
(a) Artist view of the single-molecule magnet spin-transistor presenting the one nanometer high TbPc$_{2}$ molecular magnet, contacted to the source and drain gold electrodes (yellow) and a back gate (not shown). The electronic spin of the terbium ion (pink) is coupled to the Pc ligands read-out quantum dot (white) \textit{via} exchange interaction. White lines represent electrons flow through the Pc read-out dot. This coupling modifies the chemical potential of the read-out quantum dot depending on the electronic spin state. As a result, the differential conductance is spin dependent. The orientation of the electronic spin also indicates the easy axis of the TbPc$_{2}$ molecular magnet. (b) Stability diagram of the read-out quantum dot showing the differential conductance as a function of the source-drain voltage $V_{\sf{ds}}$ and the back gate voltage $V_{\sf{g}}$. The Kondo peak on the right side of the Coulomb diamond indicates an odd number of electrons on the read-out quantum dot. Measurements of figure 2 are performed at the working point indicated by a black dot at V$_g$= -0.69V and V$_{ds}$= 0V.}
\end{center}
\end{figure}

Now, we present the characterization of our single-molecule spin transistor measuring the differential conductance as a function of the source drain voltage $V_{\sf{ds}}$ and the gate voltage $V_{\sf{g}}$ to obtain the stability diagram presented in Figure~\ref{fig1}{b} at an electronic temperature around 80 mK. Regions colored in red and blue exhibited respectively high and low differential conductance values. Note that this sample is the same as previously studied\cite{Vincent2012a}, the difference in conductance originating from a slight change of the molecules tunnel-coupling to the metallic leads due to aging of the device. From the general characteristics of the Coulomb diamond in Figure \ref{fig1}{b}, we obtained a conversion factor $\alpha=\Delta V_{\sf{ds}}/\Delta V_{\sf{g}}\approx 1/8$, resulting in a low estimation of the charging energy $E_{\sf{C}}\approx 100meV$ of the quantum dot under investigation. First, this large value agree with the idea that the single TbPc$_{2}$ molecular magnet creates the quantum dot. Moreover, Zhu \textit{et al.}~\cite{Zhu2004} showed that electrons add to the TbPc$_2$ only go to the Pc ligands up to the fifth reduction and second oxidation. As a results, as observed in our previous works on two different samples \cite{Vincent2012a,Thiele2014}, the charge state as well as the magnetic properties of the Tb$^{3+}$ ion are conserved. On these accounts, the read-out quantum dot is most likely created by the Pc ligands as depicted in Figure \ref{fig1}{a}.

\begin{figure}
\begin{center}
\includegraphics[width=0.4\textwidth]{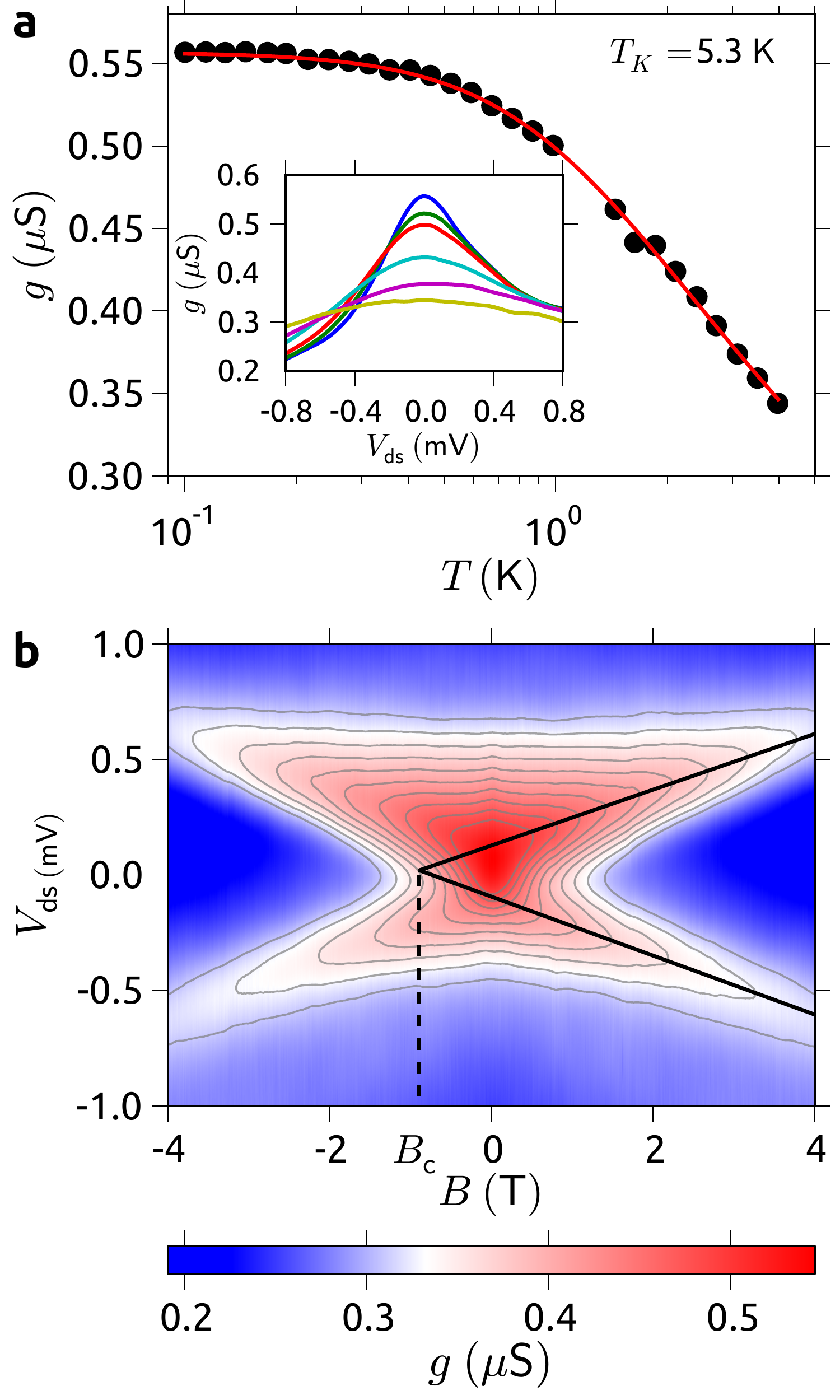}
\caption{\label{fig2}
(a) Temperature dependence of the differential conductance $\delta I/\delta V$ at $V_{\sf{ds}}=0$~V. Inset: Evolution of $\delta I/\delta V$ versus $V_{\sf{ds}}$ for a different set of temperatures. The temperatures from highest to lowest conductance were 0.1~K, 0.7~K, 1~K, 2~K, 3~K and 4.2~K. (b) Differential conductance as a function of the magnetic field $B$ and the source-drain voltage $V_{\sf{ds}}$. The solid lines were fitted to the maxima and extrapolated to negative magnetic fields. It indicates that the quantum dot under investigations possess a spin $S=1/2$ and a $g$-factor $g=2.15\pm 0.1$. The critical field, given by the intersection of the two lines, was determined as $B_{\sf{c}}=-0.88$~T, indicating a strong ferromagnetic coupling $a = -3.91$~T between the read-out quantum dot and the terbium spin.
}
\end{center}
\end{figure}

The stability diagram presented in Figure \ref{fig1}b exhibits a zero bias anomaly on the right side of the charge degeneracy point, which is associated with the usual spin S=1/2 Kondo effect observed in 2D electron gas quantum dots\cite{Goldhaber-Gordon1998a,Cronenwett1998} and single molecule transistors\cite{Park2002,Liang2002}. The Kondo temperature $T_{\sf{K}}$ was determined by measuring the differential conductance at $V_{\sf{ds}}=0$~V as a function of the temperature $T$ for a fixed gate voltage $V_{\sf{g}} = -0.69 V$ (Figure~\ref{fig2}a). By fitting the results to the empirical formula~\cite{Goldhaber-Gordon1998a}:
\begin{equation}
\label{eq1}
g(T)=g_0\left(\frac{T^2}{T^2_{\sf{K}}}\left(2^{1/s}-1\right)+1\right)^{-s}+g_{\sf{c}}
\end{equation}
where $g_0$ is the maximum conductance, $s$=0.22 and $g_{\sf{c}}$ is the fixed background conductance, we obtained a Kondo temperature $T_{\sf{K}}=5.3\pm 0.05$~K.

To determine the configuration and value of the exchange coupling between the electronic spin and the read-out quantum dot carried by the Tb$^{3+}$ ion, we investigated the evolution of the Kondo peak depending on the applied bias voltage $V_{\sf{ds}}$ and the magnetic field $B$. By increasing $B$ the Kondo peak splits linearely, with a 124~$\mu$V/T rate, as presented in Figure~\ref{fig2}b. The slope is a direct measurement of the $g$-factor $=2.15\pm 0.1$, which is consistent with the usual spin S=1/2 Kondo effect.

However, an extrapolation of this linear splitting from positive to negative magnetic fields display an intersection at a negative critical magnetic field $B_{\sf{c}}\approx-880$~mT, in contrast with the classical spin S=1/2 Kondo effect behavior. Indeed, classically $B_{\sf{c}}$ is positive and directly related to the Kondo temperature $T_{\sf{K}}$ \textit{via}: $2g\mu_{\sf{B}}B_{\sf{c}}=k_{\sf{B}}T_{\sf{K}}$. To understand this behavior, we use the analogy to the underscreened spin S=1 Kondo effect~\cite{Blandin1980,Roch2009}, where electrons of leads and a screened spin S=1/2 are  antiferromagnetically coupled. The remaining unscreened spin S=1/2 is weakened coupled by a ferromagnetic coupling to the electrons of the leads, creating an additional effective magnetic field. As a result, the critical field of finite values is decreased to almost zero Tesla~\cite{Roch2009}.

In our single molecular magnet spin-transistor device, the negative value of $B_{\sf{c}}$ originates from a ferromagnetic coupling between the read-out quantum dot and the terbium's electronic spin carrying a magnetic moment equal to 9~$\mu_{\sf{B}}$. Taking into account this coupling, the relation between the critical field $B_{\sf{c}}$ and the Kondo temperature $T_{\sf{K}}$ can be modified to~\cite{Vincent2012a}:
\begin{eqnarray}
2g\mu_{\sf{B}}B_{\sf{c}}=k_{\sf{B}}T_{\sf{K}} + a\ \mu_{\sf{B}} J_{\sf{z}} S_{\sf{z}}
\label{eq:EXC}
\end{eqnarray}
where $a$ is the coupling constant, $J_z$ and $S_z$ the z component of the electronic Tb$^{3+}$ and read-out quantum dot spins respectively. Using the Kondo temperature $T_{\sf{K}}=5.3$~K obtained from Eq.\ref{eq1} and the critical field extracted from the magnetic field dependence (Figure~\ref{fig2}b), a coupling constant $a = -3.91$~T is obtained. We emphasize that such a high value cannot be explained by a purely dipolar interaction due to the terbium magnetic moment. Indeed,  the relative distance between the phthalocyanine read-out quantum dot and the Tb$^{3+}$ ion is about  0.5~nm, giving a dipolar interaction of the order of 0.1~T, which is more than one order of magnitude lower than the measured coupling constant. As an efficient exchange interaction requires an overlap of the wave functions between the electronic magnetic moment carried by the Tb$^{3+}$ ion and the read-out quantum dot, this high coupling further validates the expected configuration for which the read-out quantum dot is the phthalocyanine. We present in the supplementary information two other TbPc$_{2}$ based spin transistors for which the exchange coupling was measured.

\begin{figure}
\begin{center}
\includegraphics[width=0.45\textwidth]{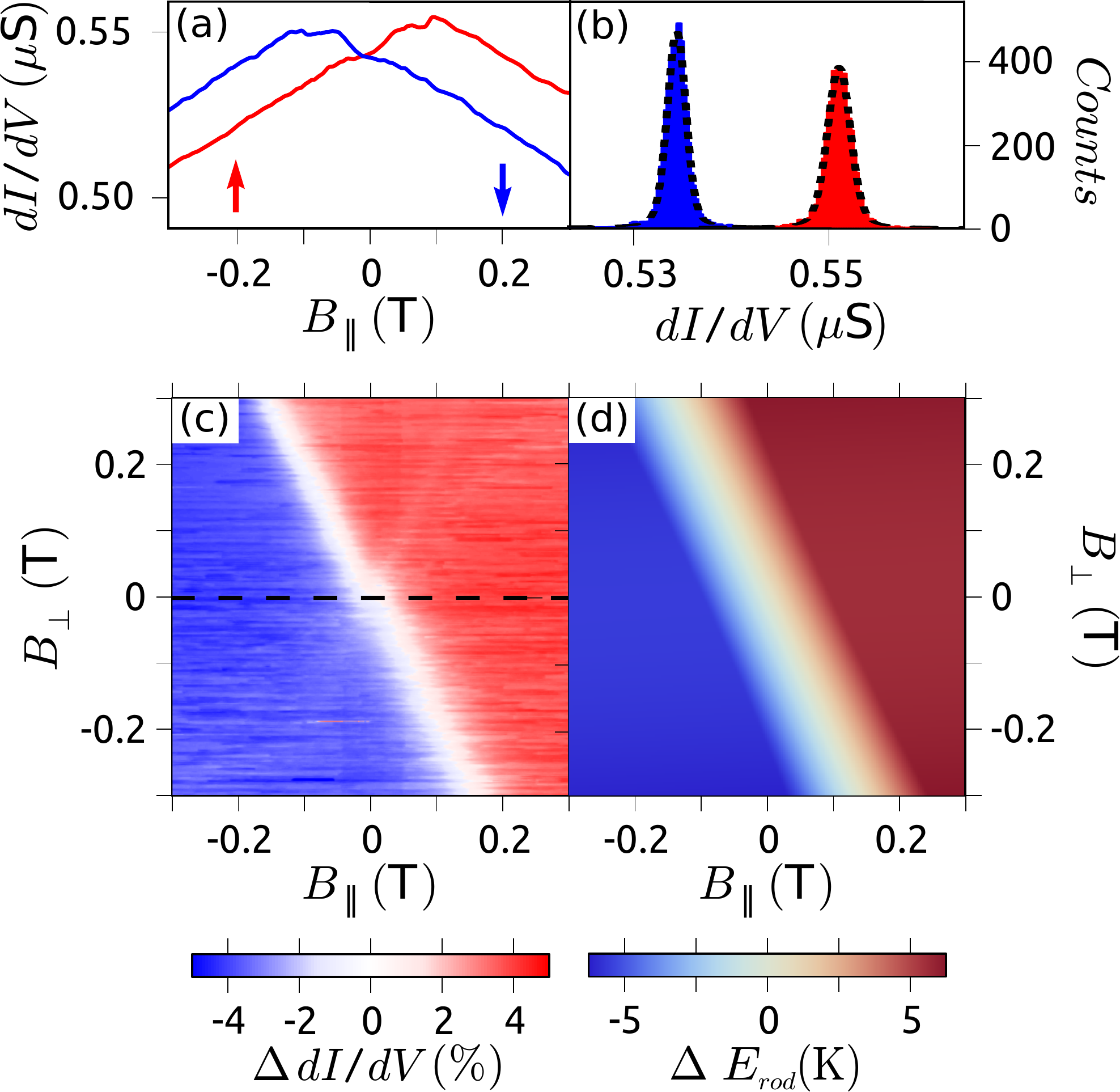}
\caption{\label{fig3}
(a) Differential conductance of the read-out quantum dot at zero bias as a function of the magnetic field $B_\parallel$ parallel to the easy axis of the molecule. The exchange coupling between the read-out quantum dot and the electronic spin results in a different conductance signal for the two spin orientations $\uparrow$ (red) and $\downarrow$ (blue). (b) Histogram of differential conductance values at $B_\parallel=+100$~mT for 10000 sweeps, showing two Gaussian-like distributions. From the overlap of the distributions we estimated a read-out fidelity of 99.5\%. (c) Conductance difference between the spin up and down as a function of the magnetic field parallel ($B_{\parallel}$) and perpendicular ($B_{\perp}$) to the easy axis of the molecule. The dotted line corresponds to the configuration of (a).  (d) Calculated energy difference $\Delta E_{rod} = E^{\uparrow}_{rod}-E^{\downarrow}_{rod}$ between the read-out quantum dot and the source and drain potential as a function of the magnetic field $B_{\parallel}$ and $B_{\perp}$.
}
\end{center}
\end{figure}

We now present the measurements and the model to explain how the exchange coupling between the electronic spin state and the read-out quantum dot induces a spin dependence of the differential conductance. We first define $B_\parallel$ and $B_{\perp}$ being the magnetic fields applied parallel and perpendicular to the easy axis of the molecule respectively (Figure \ref{fig1}{a}). For $V_{\sf{ds}}=0$ and $B_{\perp}=0$, we recorded the differential conductance at the working point while sweeping $B_\parallel$ (Figure \ref{fig3}{a}). By repeating this measurement, we obtained two distinct magneto-conductance signals, corresponding to the two electronic spin states $|\uparrow\rangle$ (red) and $|\downarrow\rangle$ (blue). The two measurements intersect at $B_\parallel=0$ and have a constant differential conductance difference for $B_\parallel>\pm100$~mT. To quantify the read-out fidelity of our device, we recorded the conductance values at $B_\parallel=100$~mT for 10000 measurements. Plotting the results into a histogram yielded two distinct Gaussian like distributions as presented in Figure~\ref{fig3}b. The read-out fidelity was determined to 99.5\% by relating the overlap of the best fits to this two distributions.

To further characterize the signal originating from the electronic spin, we determined the conductance difference between the two orientation of the electronic spin ($|\uparrow\rangle$ and $|\downarrow\rangle$) as a function of $B_\parallel$ and $B_{\perp}$(Figure \ref{fig3}c). Two different areas are visible. The red one, in which the spin $|\downarrow\rangle$ conductance was lower than the spin $|\uparrow\rangle$ and the blue one in which the inverse scenario occur. At a particular combination of $B_\parallel$ and $B_{\perp}$ the signal goes to zero, which is indicated by the white region. The configuration of Figure \ref{fig3}a is represented by the dotted line.

\section{Theoretical discussion}

To explain the magneto-conductance evolution as a function of $B_\parallel$ and $B_{\perp}$, we use  a semi-classical approach to describe the influence of the electronic spin $\boldsymbol{J}$ on the energy of the read-out quantum dot. The model considers the spin $\boldsymbol{s}$ of the read-out quantum dot, the Ising spin $\boldsymbol{J}$ of the 4f electrons both coupled to an external magnetic field $\boldsymbol{B}$. The spin $\boldsymbol{s}$ is exchanged coupled to $\boldsymbol{J}$. The Hamiltonian $\mathcal{H}$ of the system is given by:

\begin{eqnarray*}
	\label{H1}
	\mathcal{H} & = & \mathcal{H}_J +\mathcal{H}_s + \mathcal{H}_{J,s}  \\
	&=& \mathcal{H}_J +  \mathcal{H}_{rod} \\
	&=&	\mu_{\sf{B}} \boldsymbol{B} . g_{J} . \boldsymbol{J} + 	\mu_{\sf{B}} \boldsymbol{B} . \bar{\bar{g_{s}}} . \boldsymbol{s} +
	\boldsymbol{J} . \bar{\bar{a}} . \boldsymbol{s} 
\end{eqnarray*}

with $\bar{\bar{g_{s}}}$ and $g_J$ respectively the $g$-factor of the read-out quantum dot and of the electronic spin, $\bar{\bar{a}}$ the exchange coupling and $\mu_{\sf{B}}$ the Bohr magneton. In the experiment, the magnetic field $\boldsymbol{B}$ is applied along two directions, such that we define it in the x-z plane : ($\boldsymbol{B}=(B_{\perp},0,B_\parallel)$. Furthermore, $\boldsymbol{J} = \pm 6 \, \boldsymbol{e_z}$ is considered as a classical vector confined on the easy axis of the TbPc$_{2}$ molecular magnet \cite{Ishikawa2003}. Due to the axial symmetry of the system, we consider it as invariant under a rotation in the x-y plan. The read-out dot Hamiltonian can be consequently defined in the ($\perp,\parallel $) basis as: \\

\begin{equation}
\mathcal{H}_{rod} = \mu_B 
\begin{pmatrix}
B_{\perp} \\
B_{\parallel}
\end{pmatrix} 
. \bar{\bar{g_{s}}} . 
\begin{pmatrix}
s_{\perp} \\
s_{\parallel} 
\end{pmatrix} 
+
\begin{pmatrix}
J_{\perp} \\
J_{\parallel}
\end{pmatrix} 
. \bar{\bar{a}} . 
\begin{pmatrix}
s_{\perp} \\
s_{\parallel} 
\end{pmatrix} 
\end{equation}

Because the spin $\boldsymbol{s}$ of the read-out quantum dot can not be considered as a punctual electronic momentum aligned along the easy axis of terbium magnetization, the exchange interaction can not be described by a diagonal tensor. Indeed the delocalisation of the electron in the ligand involves a multi-polar correction expressed in terms of coupling between the various spacial components \textit{i.e} off-diagonal terms in the exchange tensor $\bar{\bar{a}}$. Moreover, the $g$-tensor is sensitive to the shape of the QD\cite{Hanson2007}, and measurements\cite{Hollosy2013,Takahashi2013} in quantum dot showed a significant anisotropy of the $g$-factor which turned out to be tunable by electrical means\cite{Takahashi2013,Deacon2011}.  Therefore, due to the non-symetric coupling of the read-out quantum dot to the leads, and because no chemical environment argument can ensure an isotropic $g$-factor, the more general way to express the exchange tensor $\bar{\bar{a}}$ and the $g$-factor in the ($\perp,\parallel $) basis is : \\ 

\begin{equation}
\bar{\bar{a}} = 
\begin{pmatrix}
a &  \delta a_{\parallel} \\
\delta a_{\parallel}& a 
\end{pmatrix}
\qquad
\bar{\bar{g_{s}}}= 
\begin{pmatrix}
{g_s} + \delta {g_s} & 0  \\
0 & {g_s}
\end{pmatrix}
\end{equation}

Where the notation "$\delta$" is used for the anisotropic contributions. Subsequently, taking $s_{\perp} = \hbar \sigma_x/2$ and $s_{\parallel} = \hbar \sigma_z/2$, the Hamiltonian $\mathcal{H}_{rod}$ in the read-out dot electronic spin basis is given by:

\begin{equation}
\mathcal{H}_{rod} = \frac{\hbar \, \mu_B \, g_s \, B_{\parallel}}{2} 
\begin{pmatrix}
1 & (1 + \frac{\delta {g_s}}{g_s}) \, \frac{B_{\perp}}{ B_{\parallel}}\\
(1 + \frac{\delta {g_s}}{g_s}) \, \frac{B_{\perp}}{B_{\parallel}}  & - 1
\end{pmatrix} 
 + \frac{\hbar \, a \, J_z}{2} 
\begin{pmatrix}
	 1 &  \frac{\delta a_{\parallel}}{a}\\
	 \frac{\delta a_{\parallel}}{a} & - 1 
\end{pmatrix} 
\end{equation}

The eigenenergies of the read-out dot are 
\begin{equation}
E_{J=\pm 6} = \pm [ \epsilon_0 + 2 a g_s \mu_B B_{\parallel} J_z + 2 \delta a ({g_s} + \delta {g_s}) \mu_B B_{\perp} J_z ] ^{1/2}
\end{equation}
where $\epsilon_0$ is function of $J_z^2$, meaning that the states $J_z =\pm6$ are degenerated for $ a g_s B_{\parallel} = \delta a({g_s}+\delta{g_s})B_{\perp}$. This result in a shift of the crossing point in $ B_{\parallel}$ as the function of $B_{\perp}$ observed in the measurement presented in Figure~\ref{fig3}c, given by :
\begin{equation}
B_{\parallel}^{shift} = 
\frac{({g_s} + \delta {g_s})}{g_s}
\frac{\delta a_{\parallel}}{a} 
B_{\perp}
\end{equation}

\begin{figure}
	\begin{center}
		\includegraphics[width=0.45\textwidth]{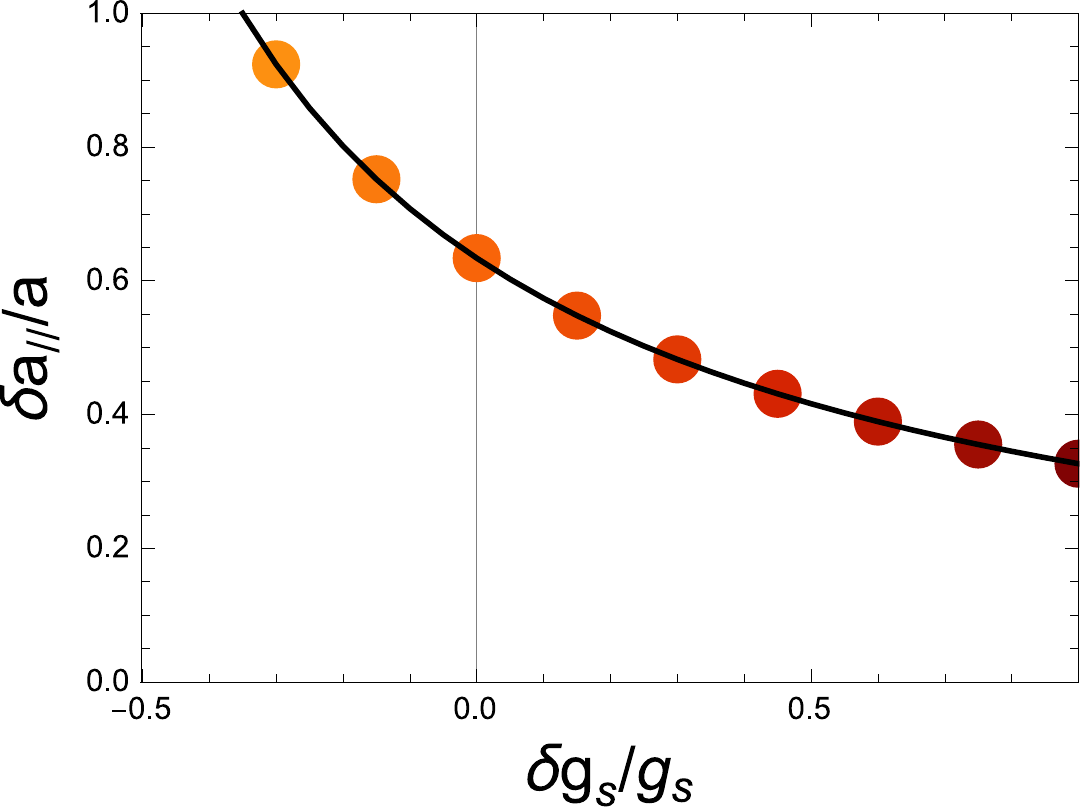}
		\caption{\label{fig4} 
			Fitting parameter line as a function of $( \frac{\delta {g_s}}{g_s} ; \frac{\delta a_{\parallel}}{a})$.}
	\end{center}
\end{figure}

In order to obtain an estimation of the off-diagonal term $\delta a_{\parallel}$, as well as the anisotropy of $g_s$, we use the experimental values determined from the measurements presented in Figure \ref{fig2} ($a = -3.91$~T and $g\approx{2.15}$), and extract the slope of  $B_{\parallel}^{shift} = -0.6 B_{\perp}$ from the measurement presented in Figure~\ref{fig3}c.  We calculated and present in Figure \ref{fig4} the different doublet $( \frac{\delta {g_s}}{g_s} ; \frac{\delta a_{\parallel}}{a})$ in accordance with the experimental measurements. An infinite number of doublet gives a perfect agreement with the experiment. Minimizing the anisotropy of the $g$-factor, we use $( \frac{\delta {g_s}}{g_s} = 0 ; \frac{\delta a_{\parallel}}{a} = 0.6 )$, we obtain the energy difference  of the read-out quantum dot : $\Delta E_{rod} = E^{\uparrow}_{rod}-E^{\downarrow}_{rod}$, depending on the state $|\uparrow\rangle$ or $|\downarrow\rangle$ of the electronic spin, as a function of $B_{\perp}$ and $B_{\parallel}$ (Figure~\ref{fig3}d). The zero sensitivity region (in white in Figure \ref{fig3}{c}) as well as the qualitative agreement comfort the model used to interpret the magneto-conductance signal.

\section{Conclusions}

In summary, we report on the proposition, theoretical explanation and experimental realization of an electrical read-out of a single electronic spin using an exchange coupled read-out quantum dot. This experimental realization has been demonstrated using the net magnetic moment of a single molecule, the read-out quantum dot being directly sensitive to the spin orientation resulting in signal amplitudes up to 4\% and read-out fidelities of 99.5\%. This detection scheme is fully consistent with any single molecule architecture for which the magnetic moment is carried by a single atom embedded by a ligand, as far as the charge state of the spin dot remains unchanged, as it is the case for all the Lanthanide Double-Decker family (Tb, Dy, Ho, \textit{etc} ... ), and could also allow the detection of a single magnetic impurity in semiconductor quantum dots, or single spin coupled to a nanotube or nanowire, leading to a potential progress in nano spintronic and quantum information processing.

\section{Methods}

The single-molecule spin based transistor was prepared by blow-drying a dilute dichloromethane solution of the TbPc$_2$ molecule onto a gold nanowire on an Au/HfO$_2$ gate fabricated through atomic-layer deposition. Before the solution was blow-dried, the electrodes were cleaned with acetone, ethanol, isopropanol solution and oxygen plasma. The connected sample was enclosed in a high-frequency, low-temperature filter, consisting of a Ecosorb microwave filter, anchored to the mixing chamber of a dilution refrigerator with a base temperature of about 50~mK. The molecule-coated nanowire was then broken by electromigration, using a voltage ramp at 50 mK. Transport measurements were taken using a lock-in amplifier in a dilution refrigerator equipped with a home-made three-dimensional vector magnet, allowing us to sweep the magnetic field in three dimensions at rates up to 0.2~T~s$^{-1}$.

\begin{acknowledgement}

This work was partially supported by MoQuaS FP7-ICT-2013-10, the DFG Programs No. SPP 1459 and No. TRR 88 3Met, ANR-12-JS10-007 SINUSManip, ANR MolQuSpin. The samples were manufactured at the NANOFAB facilities of the N\'eel Institute. The authors thank E. Eyraud, Y. Deschanels, D. Lepoittevin, C. Hoarau, E. Bonet \& C. Thirion, and are grateful to B. Canals for sharing his fruitful expertise about the nature of interactions in magnetic system

\end{acknowledgement}

\begin{suppinfo}
The following files are available free of charge.
\begin{itemize}
  \item TbPc$_{2}$ spin based transistors : All measurements presented in the article have been performed on the same sample. We present in the Supporting Information two other TbPc$_{2}$ spin based transistors for which the exchange coupling was measured.
\end{itemize}

\end{suppinfo}

\bibliography{achemso-demo}

\end{document}